\documentclass[runningheads]{llncs}
\usepackage[T1]{fontenc}
\usepackage{hyperref}
\usepackage{orcidlink}
\usepackage{booktabs}
\usepackage{float}

\usepackage{graphicx}
\begin{document}

\title{From Simulation to Sustainable Urban Mobility: A Purpose-Driven Framework for Advanced EV Charging Technology Transfer}

\titlerunning{Purpose-Driven Framework for Advanced EV Charging}

\author{
Shangqing Wang\inst{1}\orcidlink{0009-0003-9406-3554} \and
Syed Irtaza Haider\inst{1} \and
Shiwei Shen\inst{1}\orcidlink{0000-0002-0639-9492} \and
Faezeh Motazedian\inst{1} \and
Rico Radeke\inst{1} \and
Laura del Rio Carazo\inst{3}\orcidlink{0000-0002-0947-3539} \and
Frank H. P. Fitzek\inst{1,2}\orcidlink{0000-0001-8469-9573}
}

\authorrunning{S. Wang et al.}

\institute{
Deutsche Telekom Chair of Communication Networks, TU Dresden, Germany
\and
Centre for Tactile Internet with Human-in-the-Loop (CeTI), TU Dresden, Germany
\and
Escuela T{\'e}cnica Superior de Ingenieros de
Telecomunicaci{\'o}n, Universidad Polit{\'e}cnica de Madrid (UPM), Spain
\email{shangqing.wang@tu-dresden.de}
}

\maketitle

\begin{abstract}
The integration of electric vehicles (EVs) into urban mobility systems offers both opportunities and challenges for cities pursuing sustainability and energy efficiency. While simulation-based approaches have demonstrated the technical feasibility of advanced charging strategies—such as smart charging, bidirectional charging, and Vehicle-to-Grid (V2G)—their real-world impact is often constrained by policy, business models, and stakeholder engagement. This paper introduces a purpose-driven technology transfer (PDTT) framework that systematically connects simulation insights with the needs of urban stakeholders. Leveraging a modular simulation architecture based on Simulation of Urban Mobility (SUMO), we evaluate the operational and sustainability impacts of advanced EV charging in complex urban scenarios, assessing indicators such as energy efficiency and CO$_2$ emissions. The PDTT framework guides the translation of simulation results into actionable pathways for deployment, policy design, and market innovation, supporting co-creation with city planners, utilities, and communities. Our findings highlight the value of simulation as a diagnostic and engagement tool, and demonstrate how a multidimensional, stakeholder-centered approach can accelerate the sustainable adoption of advanced EV charging technologies in smart city contexts.

\end{abstract}

Acronyms used in this paper are summarized in the List of Acronyms.
\section*{List of Acronyms}
\begin{tabular}{ll}
\textbf{Acronym} & \textbf{Meaning} \\
\hline
EV      & Electric Vehicle \\
V2G     & Vehicle-to-Grid \\
V2B     & Vehicle-to-Building \\
SUMO    & Simulation of Urban Mobility \\
PDTT    & Purpose-Driven Technology Transfer \\
SUMP    & Sustainable Urban Mobility Plan \\
SUMP 2025plus & Sustainable Urban Mobility Plan 2025plus (Dresden) \\
CIVITAS & Cleaner and Better Transport in Cities (EU Initiative) \\
OEM     & Original Equipment Manufacturer \\
CO$_2$  & Carbon Dioxide \\
KPIs    & key performance indicators \\ 
\end{tabular}

\section{Introduction}
The integration of electric vehicles (EVs) into urban transportation systems presents both significant challenges and transformative opportunities for cities striving to achieve sustainability goals and optimize energy usage~\cite{apata2023risks,mahmod2015amitran,wang2024a,wang2024b}. While technical solutions such as smart charging, which optimizes charging times and power based on grid conditions, and bidirectional charging are increasingly available~\cite{sadeghian2022comprehensive}, their real-world impact is often limited by factors beyond engineering, such as business model viability, stakeholder alignment, and the complexity of urban ecosystems~\cite{wang2024b}. Bidirectional charging enables not only the charging of EVs but also the ability for EVs to discharge electricity back to the grid or other systems. This capability is commonly referred to as Vehicle-to-Grid (V2G), where EVs act as distributed energy resources, supporting grid stability and flexibility~\cite{wang2024b,mojumder2022electric}.

Recent simulation-based studies, including our previous work in Dresden's Ostra District as part of the Mobilities for EU project, have demonstrated the technical feasibility of optimizing EV charging and grid integration using tools like Simulation of Urban MObility (SUMO)~\cite{rehman2019cloud,wang2024b}. However, simulation alone is not the end goal. Instead, it serves as a \textbf{teaser}, a diagnostic tool that helps stakeholders visualize scenarios, anticipate challenges, and identify priorities for real-world implementation. The true value of simulation lies in its ability to reveal the interplay between technical, economic, social, and regulatory dimensions that shape the success of EV integration.

Building on the simulation architecture described in our conference paper~\cite{Wang2025:Simulationb}, this extended work introduces a \textbf{purpose-driven technology transfer (PDTT) framework}. This framework synthesizes insights from simulation, business model development, and stakeholder engagement to bridge the gap between engineering innovation and practical adoption. The PDTT framework is illustrated through the Dresden case, with a focus on how simulation results can inform policy design, market incentives, infrastructure planning, and social readiness.

The transition from simulated energy optimization models to real-world deployment is complex and requires more than technical validation. While this paper does not present results from a full-scale operational pilot or detailed hardware deployment, it extends our previous simulation work by demonstrating how advanced charging strategies, specifically, smart charging and bidirectional (V2G) charging, can be modeled and evaluated in a realistic urban context. We further detail how the PDTT framework enables a multidimensional, stakeholder-centered approach to sustainable EV technology transfer and discuss its relevance for broader urban mobility initiatives.

\section{Background and Related Work}
\subsection{Simulation Architectures for EV Integration}
Simulation environments have become essential tools for modeling the complex interactions among EVs, charging infrastructure, and urban power grids. The SUMO platform, in particular, enables detailed modeling of traffic patterns, vehicle movements, and charging behaviors, providing a foundation for scenario-based analysis and optimization~\cite{krajzewicz2012recent,kurczveil2014implementation}. Recent work has integrated SUMO with custom energy management systems and the Amitran methodology to evaluate the sustainability impacts of various charging strategies, including CO$_2$ emissions and grid stability~\cite{mahmod2015amitran,Wang2025:Simulationb}.

Our approach builds on this foundation by developing a modular simulation architecture that supports the assessment of operational and sustainability impacts for advanced EV charging strategies, such as smart charging and bidirectional (V2G) charging, in complex urban scenarios. The architecture is designed to be flexible and extensible, allowing for the integration of established methods for quantifying environmental impacts, including CO$_2$ emissions, as the project evolves.

Co-simulation frameworks further enhance the ability to analyze complex systems by coupling multiple simulation units and facilitating real-time data exchange~\cite{rehman2019cloud,rohjans2013mosaik,barbierato2022comparison}. These approaches enable city planners and fleet managers to explore "what-if" scenarios, assess the implications of different strategies, and make informed decisions regarding EV integration~\cite{rehman2019cloud,steinbrink2018smart}.

\subsection{Challenges Beyond Simulation: Business, Stakeholders, and Sustainability}

Despite advances in simulation and technical optimization, the deployment of V2G and smart charging technologies in real urban environments faces persistent challenges. These include infrastructure limitations, regulatory uncertainty, business model complexity, and the need for stakeholder alignment~\cite{apata2023risks,Wang2025:Simulationb}. Prior research highlights that successful EV integration requires not only robust technical solutions but also frameworks that address social acceptance, value distribution, and policy engagement~\cite{wang2014evaluation,barbierato2022comparison}.

The Mobilities for EU project exemplifies this multidimensional challenge, aiming to enhance urban mobility through the integration of EVs, smart charging, and advanced traffic management, while aligning with broader sustainability goals~\cite{MobilitiesForEU,barbierato2022comparison}. Our previous work in Dresden’s Ostra District demonstrated that simulation can provide valuable insights for planners, but also revealed the limitations of technical approaches when not coupled with business and stakeholder considerations.

\subsection{Toward Purpose-Driven Technology Transfer}
Emerging literature and recent EU project experiences underscore the necessity of frameworks that systematically link simulation outputs with actionable pathways for technology transfer and sustainable adoption~\cite{christensen2020agent,trevena2022modeling,collste2017policy}. This includes the development of business models tailored to advanced EV charging, stakeholder engagement processes, and metrics for both technical and societal impact~\cite{mahmod2015amitran,rehman2019cloud,barbierato2022comparison}. Drawing on these insights, this paper proposes a comprehensive, purpose-driven framework for sustainable Advanced EV Charging technology transfer, extending beyond simulation to address real-world impact and adoption. 

\section{Simulation Architecture Recap}

\begin{figure}[ht]
    \centering
    \includegraphics[width=\columnwidth]{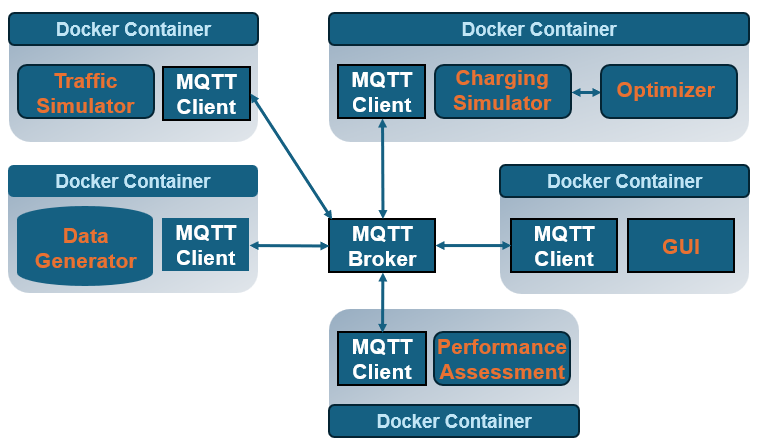}
    \caption{Simulation architecture developed for the Mobilities for EU project in Dresden’s Ostra District. 
Reproduced with permission from~\cite{Wang2025:Simulationb}. © 2025 SCITEPRESS.}
    \label{fig:sim-arch}
\end{figure}

Figure~\ref{fig:sim-arch}, reproduced from our earlier conference paper~\cite{Wang2025:Simulationb}, 
shows the simulation architecture developed for the Mobilities for EU project in Dresden’s Ostra District. This architecture was designed to optimize EV charging strategies and assess their sustainability impacts in a real urban context~\cite{Wang2025:Simulationb}. Leveraging SUMO platform, the architecture models detailed traffic patterns and vehicle movements, while a custom energy management system enables both smart and bidirectional (V2G) charging~\cite{krajzewicz2012recent,kurczveil2014implementation}.

Our simulation environment enables the assessment of Carbon emissions and energy efficiency through well-established methods, including well-to-tank analysis, to capture both direct and indirect impacts of different charging strategies~\cite{mahmod2015amitran}. While we acknowledge the Amitran methodology as a systematic approach for such evaluations, the current focus is on developing a flexible, modular architecture that can incorporate various sustainability assessment methods as the project evolves.

The modular, scalable simulation environment supports a range of scenarios, such as smart charging, peak shaving, and V2G services. Co-simulation allows synchronized data exchange between traffic and energy domains, enabling dynamic, real-time adjustments to charging strategies~\cite{rehman2019cloud,rohjans2013mosaik}.

\subsection{Key Technical Findings from the Simulation}
Our previous work demonstrated several important benefits of smart and bidirectional charging strategies compared to conventional approaches:

\begin{itemize}
    \item \textbf{Peak load reduction and energy efficiency}: Smart charging and V2G strategies can significantly reduce peak grid loads and improve energy efficiency in urban districts.
    \item \textbf{Potential for CO\textsubscript{2} reduction}: Scenario analysis indicates that meaningful emissions reductions are possible, particularly when charging aligns with renewable energy availability. While our current simulation architecture is designed to support systematic CO$_2$ assessment using established methods, detailed emissions calculations will be conducted in future work as the framework is further developed and validated.
    \item \textbf{Modularity and adaptability:} The architecture’s modular design allows easy adaptation to different city sizes, EV penetration rates, and infrastructure configurations.
\end{itemize}

\subsection{Illustrative Results}
\begin{figure}[!ht]
\centering
\includegraphics[width=\columnwidth]{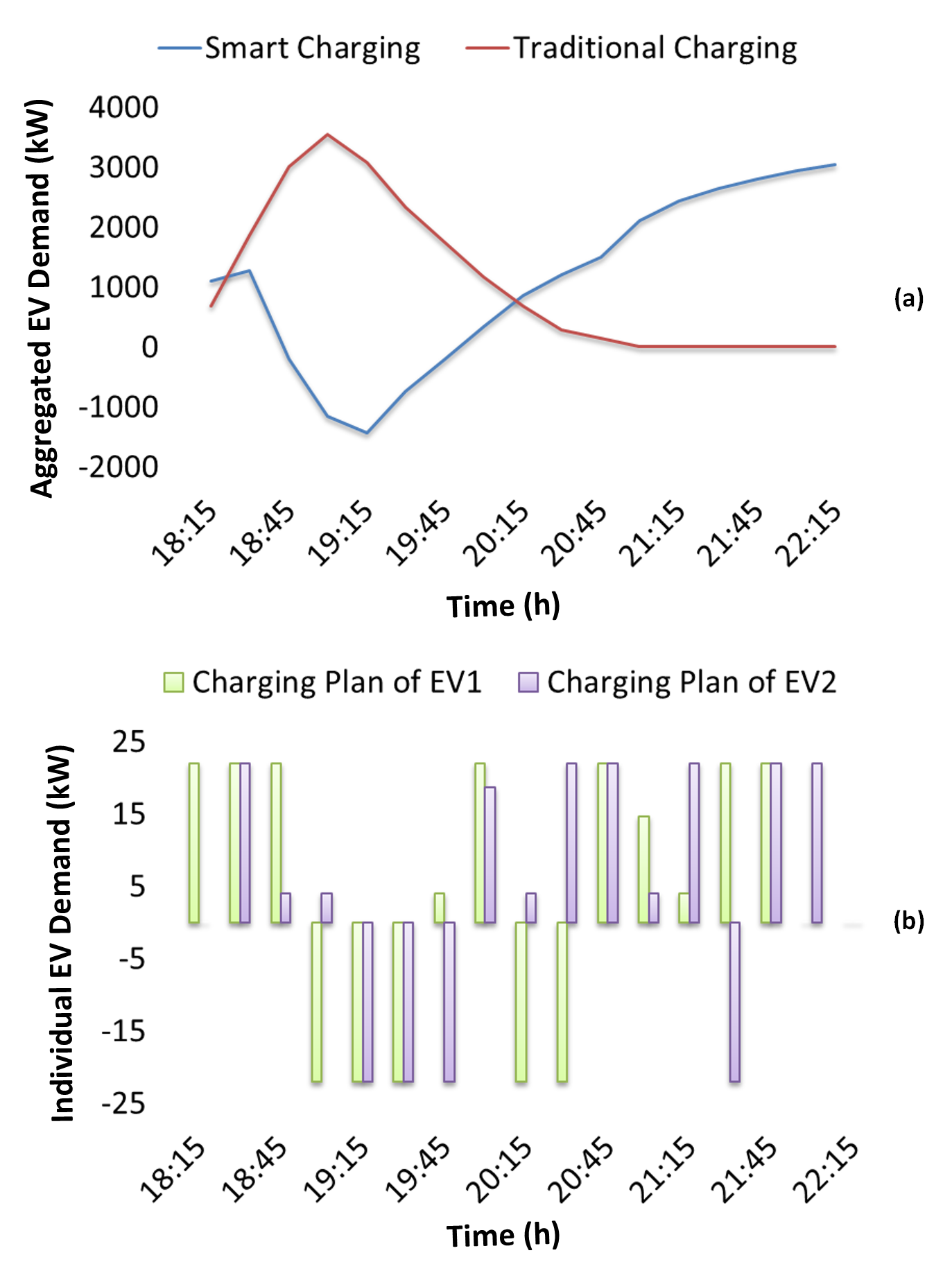}
\caption{Aggregated EV charging demand: traditional vs. smart charging. 
Adapted from~\cite{Wang2025:Simulationb}. © 2025 SCITEPRESS.}
\label{fig:smartchargin}
\end{figure}

To visualize these impacts, Figure~\ref{fig:smartchargin}, adapted from~\cite{Wang2025:Simulationb}, compares the aggregated EV charging demand under traditional and smart charging scenarios. The results highlight the ability of smart charging to flatten demand peaks and shift consumption to off-peak periods, supporting both grid stability and cost efficiency.

\begin{figure}[!ht]
\centering
\includegraphics[width=\columnwidth]{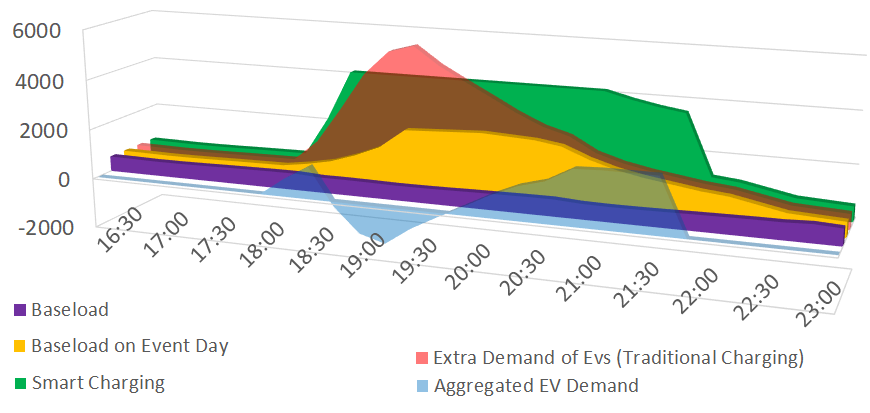}
\caption{Comparison of stadium baseload, event day baseload, uncoordinated EV charging demand, 
and peak shaving using smart charging. Adapted from~\cite{Wang2025:Simulationb}. © 2025 SCITEPRESS.}
\label{fig:stadium-baseload}
\end{figure}

Figure~\ref{fig:stadium-baseload}, adapted from our prior work~\cite{Wang2025:Simulationb}, presents the stadium district baseload with and without coordinated EV charging, demonstrating further opportunities for peak shaving and system optimization.

Table~\ref{tab:electricity_prices}, adapted from~\cite{Wang2025:Simulationb},  illustrates how dynamic charging can respond to fluctuations in electricity prices, further enhancing economic efficiency.

\begin{table}[h!]
    \centering
    \caption{Electricity Price for the period between 18:00 to 22:00
    Adapted from~\cite{Wang2025:Simulationb}. © 2025 SCITEPRESS.}
    \label{tab:electricity_prices}
    \begin{tabular}{lccc}
        \toprule
        \textbf{Category} & \textbf{Price(€/kWh)} \\
        \midrule
        Minimum Electricity Price & 0.064 \\
        Mean Electricity Price & 0.172 \\
        Maximum Electricity Price & 0.264 \\
        Average Cost with Traditional Charging & 0.184 \\
        Average Cost with Smart Charging & 0.170 \\
        \bottomrule
    \end{tabular}
\end{table}

\subsection{Impact on City Planning and Stakeholder Engagement}
The simulation results provided Dresden’s city planners and stakeholders with a data-driven preview of the operational and sustainability benefits of deploying smart and bidirectional charging solutions. Beyond technical feasibility, the simulation architecture has already proven valuable as a communication and decision-support tool, catalyzing engagement with city authorities and utility partners, supporting the co-definition of sustainability key performance indicators (KPIs), and informing the design of future business models and policy strategies.

While simulation provides the technical evidence and stakeholder “teaser” needed to spark interest and inform decision-making, it is only the starting point. The Real-world deployment of advanced EV charging solutions requires a systematic approach that addresses technical feasibility, business viability, regulatory alignment, and stakeholder engagement. To meet this need, we introduce the PDTT framework, a practical, multidimensional roadmap that leverages simulation insights to de-risk V2G deployment and keep sustainability as the guiding purpose.

\section{Theoretical Foundations: From Purpose and Dimensions to Action}
The successful transfer and scaling of advanced EV charging technologies in urban environments requires more than technical innovation—it demands a holistic, purpose-driven approach that integrates business, regulatory, stakeholder, and technological dimensions from the outset. Our framework for sustainable technology transfer is grounded in two complementary strands of recent research and practice: the multidimensional project management approach of del Río Carazo et al.~\cite{delRioCarazo2022} and the Purpose Code philosophy introduced by Grumet~\cite{grumet2025purpose}.

Del Río Carazo et al.~\cite{delRioCarazo2022} emphasize that effective technology transfer is not achieved through isolated technical solutions, but through the coordinated integration of four key dimensions: Regulation (R), Business Models (BM), Stakeholder Engagement (ST), and Technology/Digitalization (Tech Dig). Their framework, originally developed for sustainable rural electrification, demonstrates that these elements must be addressed iteratively and in concert throughout all project phases—from preparation and co-design to implementation and evaluation. This approach ensures that technical, economic, social, and regulatory factors are not only considered together, but that stakeholder participation and adaptability are embedded at every stage. While developed in a rural context, this multidimensional perspective is highly relevant for complex urban mobility projects, where similar cross-cutting challenges and opportunities arise.

The Purpose Code framework~\cite{grumet2025purpose} further shapes our approach by placing a single, overarching “Big P”—Purpose—at the center of all action. In our context, this Purpose is defined as \textbf{sustainability}: maximizing positive environmental, social, and economic impact through urban mobility innovation. The Purpose Code also advocates for “little Ps”—the key dimensions (such as business, technology, regulation, and stakeholders)—that operationalize the central Purpose in practice.

By integrating these foundations, our framework adopts sustainability as the guiding Purpose (Big P), realized through the four strategic dimensions (little Ps) identified by del Río Carazo et al. We then translate these dimensions into a practical, actionable process via a seven-step roadmap. Each step draws on one or more of the four core elements, ensuring the framework is both multidimensional and grounded in real-world project management best practices.

\begin{quote}
In summary, our framework is structured as follows:
\textit{Sustainability} is the core Purpose (big P), realized through four foundational dimensions—Business Models, Regulation, Stakeholder Engagement, and Technology/Digitalization (little Ps)—as articulated by del Río Carazo et al. These dimensions are operationalized through a seven-step process, providing clear, actionable guidance for the technology transfer of advanced EV charging solutions in urban mobility.
\end{quote}

\begin{figure}[ht]
\centering
\includegraphics[width=\linewidth, trim=1.5cm 0.5cm 1.5cm 0.5cm, clip]{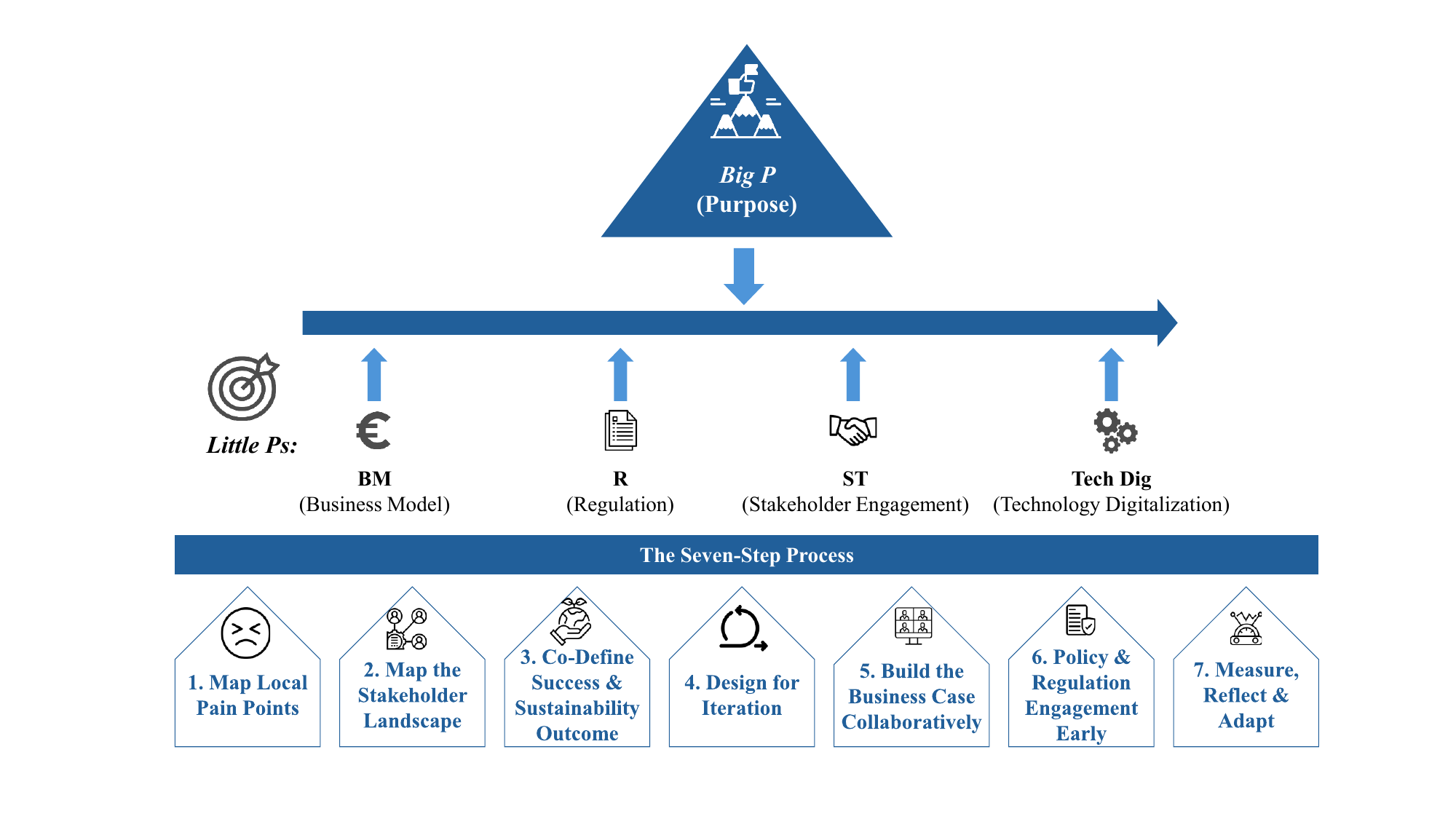}
\caption{Purpose-driven technology transfer framework for advanced EV charging. Sustainability (\textit{Big P}), represented by the triangle at the top, serves as the overarching goal. Four foundational dimensions (\textit{Little Ps}): Business Models (BM), Regulation (R), Stakeholder Engagement (ST), and Technology \& Digitalization (Tech Dig), form the structural base. A seven-step, bottom-up process operationalizes these dimensions, guiding projects from identifying local needs to defining success metrics, stakeholder engagement, iterative design, policy alignment, and business case development. The framework is tailored for advanced EV charging technology transfer, ensuring that technical innovation aligns with systemic sustainability impact.}
\label{fig:purpose-framework}
\end{figure}

Figure~\ref{fig:purpose-framework} visually represents this integration, showing how the seven steps connect the strategic dimensions to the overarching purpose of sustainability.

\section{Purpose-Driven 7-Step Framework for Advanced EV Charging Technology Transfer}
\label{sec:purpose-7-steps}

Simulation results, such as lower energy costs, peak shaving, and projected CO$_2$ reductions, not only demonstrate the technical potential of advanced EV charging solutions but also serve as powerful engagement tools for stakeholders. However, to move from simulation to sustainable, real-world deployment, a structured process is needed. The PDTT framework provides this process, systematically translating simulation insights into actionable steps for technology transfer, stakeholder alignment, and business model innovation, all under the guiding purpose of sustainability.

Building directly on the theoretical foundations outlined above—including the multidimensional project management approach of del Río Carazo et al.~\cite{delRioCarazo2022} and the Purpose Code philosophy~\cite{grumet2025purpose}—we present a practical, purpose-driven framework to guide advanced EV Charging technology transfer in urban mobility. At its core, sustainability serves as the guiding purpose, operationalized through four foundational dimensions: Business Models, Regulation, Stakeholder Engagement, and Technology/Digitalization.

Our framework translates these principles into a clear, actionable seven-step process. Each step is designed to address one or more of the foundational dimensions, reflecting the complex and interconnected nature of sustainable innovation. This dual top-down (purpose- and strategy-driven) and bottom-up (data- and feedback-driven) approach ensures that technical, business, regulatory, and stakeholder considerations are integrated throughout the project lifecycle.

\subsection{The Seven Steps}

\begin{enumerate}
\item \textbf{Map Local Pain Points} (Technology/Digitalization)\\
Identify context-specific technical and operational challenges through simulation and data analysis, ensuring solutions are tailored to real-world needs.

\item \textbf{Map the Stakeholder Landscape} (Stakeholder Engagement)\\
Systematically identify and engage all relevant actors—city planners, utilities, fleet operators, and end-users—to ensure early buy-in and shared understanding.

\item \textbf{Co-Define Success and Sustainability Outcomes} (Business Models, Stakeholder Engagement, Regulation)\\
Collaboratively set clear, measurable sustainability goals and KPIs, aligning technical, business, and policy objectives with the overarching purpose.

\item \textbf{Design for Iteration—Pilots as Learning Labs} (Technology/Digitalization, Stakeholder Engagement)\\
Structure pilot projects for iterative testing, learning, and adaptation, incorporating feedback from both technical performance and stakeholder experience.

\item \textbf{Build the Business Case—Collaborative} (Business Models)\\
Develop robust, context-appropriate business models that distribute value and risk among stakeholders, informed by simulation and financial analysis.

\item \textbf{Policy and Regulatory Engagement Early} (Regulation)\\
Proactively identify and address regulatory barriers, ensuring compliance and alignment with local and national policy frameworks from the outset.

\item \textbf{Measure, Reflect, and Adapt—Purpose Over Hype} (All Dimensions)\\
Continuously monitor progress against KPIs, reflect on outcomes, and adapt strategies to maintain alignment with the central purpose of sustainability.
\end{enumerate}

The following table~\ref{tab:sim-to-pdtt} summarizes how key simulation outputs can be leveraged at each stage of the PDTT process to support robust, sustainable technology transfer.

\begin{table}[h]
\centering
\caption{Mapping of Simulation Outputs to PDTT Framework Steps}
\label{tab:sim-to-pdtt}
\begin{tabular}{ll}
\hline
\textbf{Simulation Output}         & \textbf{Informs PDTT Step(s)} \\
\hline
Peak load reduction                & Map Local Pain Points, Build the Business Case \\
Energy cost optimization           & Co-Define Success, Build the Business Case \\
Potential CO$_2$ reduction         & Co-Define Success, Policy/Regulatory Engagement \\
Stakeholder feedback     & Stakeholder Mapping, Pilot Design, Iteration \\
\hline
\end{tabular}
\end{table}

\subsection{Framework Integration and Application}
This seven-step process operationalizes our theoretical foundations, providing a structured yet flexible roadmap for cities and project teams. By systematically addressing the technical, business, regulatory, and social dimensions at each stage, the framework ensures that advanced EV charging technology transfer is not only technically robust but also socially and economically viable. The approach is adaptable to diverse urban contexts and supports iterative learning and continuous improvement.

Table~\ref{tab:seven-steps-dimensions} presents an overview of how each step of the framework aligns with the four core dimensions (Business Models, Regulation, Stakeholder Engagement, and Technology/Digitalization), highlighting the multidimensional nature of the PDTT approach.

\begin{table}[h]
\centering
\caption{Mapping of Seven Steps to Core Dimensions}
\label{tab:seven-steps-dimensions}
\begin{tabular}{ll}
\hline
\textbf{Step} & \textbf{Primary Dimension(s)} \\
\hline
Map Local Pain Points & Technology/Digitalization \\
Map the Stakeholder Landscape & Stakeholder Engagement \\
Co-Define Success and Sustainability Outcomes & BM, ST, R \\
Design for Iteration—Pilots as Learning Labs & Tech Dig, ST \\
Build the Business Case—Collaborative & Business Models \\
Policy and Regulatory Engagement Early & Regulation \\
Measure, Reflect, and Adapt & All Dimensions \\
\hline
\end{tabular}
\end{table}

\section{Business Model Dimension for Advanced EV Charging Technology Transfer}
\label{sec:business-model}
Business model innovation is a critical enabler for the successful adoption and scaling of advanced EV charging solutions—including smart charging, bidirectional charging, and V2G, in urban mobility~\cite{arena2022,bohnsack2014,kaufmann2017,wang2025bm}. Business models define how organizations create, deliver, and capture value in the evolving energy and transport landscape. For advanced EV charging, this means orchestrating value flows among vehicle owners, energy system operators, intermediaries, and transport industry stakeholders, often requiring new forms of cooperation and risk-sharing~\cite{arena2022,kaufmann2017,bohnsack2014}.

Recent studies emphasize that business models for advanced EV charging are highly context-dependent and must be tailored to local market structures, regulatory environments, and stakeholder capabilities~\cite{arena2022,bohnsack2014,steward2017}. Compared to vehicle-to-Building (V2B), V2G and bidirectional charging business models are more complex due to the need to coordinate relationships across the energy and transport sectors, manage technical integration challenges, and align incentives for all actors involved~\cite{arena2022,bohnsack2014}. For example, fleet operators can benefit from reduced total cost of ownership; energy market participants can optimize trading and balancing; and network operators can leverage V2G for grid stability and congestion management~\cite{wang2025bm,arena2022,steward2017}.

A viable business model for advanced EV charging typically addresses:
\begin{itemize}
    \item \textbf{Value Proposition:} Grid services (e.g., frequency control, peak shaving), cost savings, and sustainability benefits for each stakeholder group~\cite{arena2022,bohnsack2014}.
    \item \textbf{Customer Segments:} Fleet managers, building owners, utilities, and energy market participants~\cite{wang2025bm,bohnsack2014}.
    \item \textbf{Revenue Streams:} Payments for ancillary services, energy trading, and optimized fleet operations~\cite{arena2022,kaufmann2017,steward2017}.
    \item \textbf{Key Partnerships:} Collaboration between OEMs, charging infrastructure providers, energy retailers, and network operators~\cite{wang2025bm,kaufmann2017}.
    \item \textbf{Cost Structure:} Investment in hardware/software, operational costs, and potential grid upgrade savings~\cite{arena2022,steward2017}.
\end{itemize}

Our simulation-driven business model design in the Dresden pilot demonstrates how leveraging advanced communication technologies (e.g., 5G) and iterative prototyping enables stakeholders to quantify value, test incentive mechanisms, and identify scalable adoption pathways~\cite{wang2024a,wang2025bm}. This aligns with the literature’s emphasis on stakeholder-driven configuration and iterative refinement of business models~\cite{arena2022,bohnsack2014}. As advanced EV charging matures, business models will likely evolve to support greater convergence between energy and transport markets, more dynamic pricing, and deeper integration with building and grid management systems~\cite{arena2022,kaufmann2017}.

\section{From Simulation to Real-World Implementation: The Dresden Case}
Dresden’s Ostra Park, as part of the Mobilities for EU project ~\cite{MobilitiesForEU}, serves as a real-world testbed for operationalizing our purpose-driven, multidimensional framework, providing a unique opportunity to translate simulation insights and business model innovation into actionable urban mobility solutions. Scenario analyses generated by our simulation architecture—quantifying impacts such as peak load reduction, cost savings, and CO$_2$ emissions—directly inform the selection of pilot configurations, value propositions, and incentive mechanisms. These data-driven insights are used in workshops with city planners, utilities, and fleet operators to co-create business models that align technical feasibility with local economic and regulatory realities.

The business model framework, grounded in simulation results and iterative stakeholder consultations, aligns the interests of city authorities, utilities, fleet operators, and residents. It maps out feasible revenue streams, cost-sharing arrangements, and partnership models, ensuring that all actors have a stake in the pilot’s success. This iterative, stakeholder-driven process, central to both the CIVITAS~\cite{CIVITAS2017} and SUMP 2025plus frameworks~\cite{SUMP2025plus}, ensures that pilot planning remains flexible and responsive to evolving feedback.

Currently, Dresden’s Department of Economic Development is leading the coordination, bringing together administration, industry, and research partners to advance the pilot from concept to implementation. Regular workshops and working groups with city planners and mobility stakeholders are focused on refining technical requirements, business case details, and regulatory pathways. Key steps toward deployment include securing local approvals, finalizing technical specifications for charging infrastructure, and designing user engagement campaigns to ensure broad acceptance.

The process is inherently adaptive: simulation results are continuously updated based on stakeholder feedback and emerging field data, ensuring that both technical and business frameworks remain responsive to local needs and constraints. This approach not only de-risks the transition from simulation to real-world application but also builds local capacity for scaling up advanced EV charging solutions citywide.

By grounding pilot design and business model development in rigorous simulation and stakeholder engagement, the Dresden case provides a replicable pathway for other cities seeking to de-risk and accelerate the adoption of advanced EV charging solutions.

\section{Conclusions}
This paper introduces the PDTT framework, a novel, multidimensional approach that systematically bridges simulation insights with stakeholder needs for the sustainable deployment of advanced EV charging solutions, including smart charging, bidirectional charging, and V2G, within the context of urban mobility. By integrating multidimensional project management principles, business model innovation, and simulation-driven scenario analysis, our approach addresses the technical, economic, regulatory, and stakeholder challenges that have historically limited real-world deployment.

Our seven-step framework provides actionable guidance for cities and project teams, ensuring that advanced EV charging technologies are not only technically feasible but also socially and economically viable. The Dresden case study demonstrates how simulation insights and collaborative business modeling can inform the design and implementation of real-world pilots, de-risking the transition from research to practice.

Looking ahead, we propose the development of a comprehensive multicriteria management model aligned with the CIVITAS evaluation framework to further enhance the assessment of sustainability impacts. Future research will focus on validating this framework with empirical data from ongoing pilots, refining business models through iterative stakeholder engagement, and exploring the integration of renewable energy sources and dynamic pricing mechanisms. By advancing both theory and practice, this work aims to accelerate the sustainable adoption of advanced EV charging technologies in smart cities worldwide.

\section{Statement of Differences}
This journal article significantly extends the previously published conference version in several key areas:
\begin{itemize}
    \item \textbf{Framework Innovation:} We introduce the new PDTT framework, a seven-step, multidimensional approach for advanced EV charging technology transfer, grounded in project management best practices and the Purpose Code philosophy. This PDTT framework was not present in the conference version and represents a core conceptual advance.
    \item \textbf{Business Model Dimension:} The paper presents an expanded and detailed analysis of business model innovation, including stakeholder mapping, value proposition design, and revenue stream identification, informed by recent literature and pilot work in Dresden.
    \item \textbf{Integration with Policy and Evaluation Frameworks:} We explicitly align our approach with the CIVITAS and SUMP 2025plus frameworks, situating our work within leading European urban mobility initiatives.
    \item \textbf{Validation Plan:} The paper details a clear pathway from simulation to real-world implementation, using the Dresden Ostra Park pilot as a case study. It outlines an iterative, stakeholder-driven process for pilot deployment and ongoing validation.
    \item \textbf{Expanded Literature Review and Context:} The journal version provides a more comprehensive review of recent developments in advanced EV charging, co-simulation, and urban mobility policy.
    \item \textbf{Future Research Directions:} We propose the development of a multicriteria management model for sustainability evaluation, setting a clear agenda for future empirical validation and framework refinement.
\end{itemize}

These enhancements make the present article a substantially broader and deeper contribution, offering both a robust theoretical foundation and a practical roadmap for sustainable technology transfer in urban mobility.

\begin{credits}
\section*{Acknowledgement}
This work is funded by the European Union under Grant Agreement No 101139666, MOBILITIES FOR EU. Views and opinions expressed are those of the authors only and do not necessarily reflect those of the European Union or the European Climate, Infrastructure and Environment Executive Agency (CINEA). Neither the European Union nor the granting authority can be held responsible for them. This work is supported by the German Federal Ministry for Economic Affairs and Climate Action (BMWK) under project ID 03EI6082A, DymoBat, the German Research Foundation (DFG) as part of Germany's Excellence Strategy—EXC 2050/1—Cluster of Excellence “Centre for Tactile Internet with Human-in-the-Loop” (CeTI) of Technische Universität Dresden under project ID 390696704 and the Federal Ministry for Education and Research (BMBF) in the program of “Souverän. Digital. Vernetzt.” Joint project 6G-life, grant number 16KISK001K.

\subsubsection{\discintname}
The authors have no competing interests to declare.
\end{credits}
%
%
%
%
\clearpage
\bibliographystyle{splncs04}
\bibliography{references}

\end{document}